\documentclass[sigconf]{acmart}
\usepackage{amsmath,amsfonts}
\usepackage{bbm}
\usepackage{physics}
\usepackage{algorithmic}
\usepackage{graphicx}
\usepackage{textcomp}
\usepackage{xcolor}
\usepackage{tabu}
\usepackage{qcircuit}
\usepackage{enumitem}

\AtBeginDocument{%
  \providecommand\BibTeX{{%
    Bib\TeX}}}

\DeclareUnicodeCharacter{2009}{ }

\copyrightyear{2023} 
\acmYear{2023} 
\setcopyright{acmlicensed}
\acmConference[ISCA '23] {Proceedings of the 50th Annual International Symposium on Computer Architecture}{June 17--21, 2023}{Orlando, FL, USA.}
\acmBooktitle{Proceedings of the 50th Annual International Symposium on Computer Architecture (ISCA '23), June 17--21, 2023, Orlando, FL, USA}
\acmPrice{15.00}
\acmISBN{979-8-4007-0095-8/23/06} 
\acmDOI{10.1145/3579371.3589106}

\def\BibTeX{{\rm B\kern-.05em{\sc i\kern-.025em b}\kern-.08em
    T\kern-.1667em\lower.7ex\hbox{E}\kern-.125emX}}

\begin{document}

\title[Dancing the Quantum Waltz: Compiling Three-Qubit Gates on Four Level Architectures]{Dancing the Quantum Waltz:\\Compiling Three-Qubit Gates on Four Level Architectures}

\author{Andrew Litteken}
\email{litteken@uchicago.edu}
\orcid{0000-0001-5676-1747}
\affiliation{%
  \institution{University of Chicago}
  \city{Chicago}
  \state{Illinois}
  \country{USA}
}

\author{Lennart Maximilian Seifert}
\email{lmseifert@uchicago.edu}
\orcid{0000-0002-2605-3720}
\affiliation{%
  \institution{University of Chicago}
  \city{Chicago}
  \state{Illinois}
  \country{USA}
}

\author{Jason D. Chadwick}
\email{jchadwick@uchicago.edu}
\orcid{0000-0002-7932-1418}
\affiliation{%
 \institution{University of Chicago}
 \city{Chicago}
  \state{Illinois}
 \country{USA}
}

\author{Natalia Nottingham}
\email{nottingham@uchicago.edu}
\orcid{0000-0003-3824-074X}
\affiliation{%
  \institution{University of Chicago}
  \city{Chicago}
  \state{Illinois}
  \country{USA}
}

\author{Tanay Roy}
\email{roytanay@fnal.gov}
\orcid{0000-0003-0438-012X}
\affiliation{%
  \institution{Fermilab}
  \city{Batvia}
  \state{Illinois}
  \country{USA}
}

\author{Ziqian Li}
\email{ziqianli@stanford.edu}
\orcid{0000-0002-3419-2333}
\affiliation{%
  \institution{Stanford University}
  \city{Stanford}
  \state{California}
  \country{USA}
}

\author{David Schuster}
\email{dschus@stanford.edu}
\orcid{0000-0002-0012-3874}
\affiliation{%
  \institution{Stanford University}
  \city{Stanford}
  \state{California}
  \country{USA}
}

\author{Frederic T. Chong}
\email{chong@cs.uchicago.edu}
\orcid{0000-0001-9282-4645}
\affiliation{%
  \institution{University of Chicago}
  \city{Chicago}
  \state{Illinois}
  \country{USA}
}

\author{Jonathan M. Baker}
\email{jonathan.baker@duke.edu}
\orcid{0000-0002-0775-8274}
\affiliation{%
  \institution{Duke University}
  \city{Durham}
  \state{North Carolina}
  \country{USA}
}

\begin{abstract}
    Superconducting quantum devices are a leading technology for quantum computation, but they suffer from several challenges.  Gate errors, coherence errors and a lack of connectivity all contribute to low fidelity results. In particular, connectivity restrictions enforce a gate set that requires three-qubit gates to be decomposed into one- or two-qubit gates.  This substantially increases the number of two-qubit gates that need to be executed. However, many quantum devices have access to higher energy levels.  We can expand the qubit abstraction of $\ket{0}$ and $\ket{1}$ to a ququart which has access to the $\ket{2}$ and $\ket{3}$ state, but with shorter coherence times.  This allows for two qubits to be encoded in one ququart, enabling increased virtual connectivity between physical units from two adjacent qubits to four fully connected qubits. This connectivity scheme allows us to more efficiently execute three-qubit gates natively between two physical devices.

    We present direct-to-pulse implementations of several three-qubit gates, synthesized via optimal control, for compilation of three-qubit gates onto a superconducting-based architecture with access to four-level devices with the first experimental demonstration of four-level ququart gates designed through optimal control.  We demonstrate strategies that temporarily use higher level states to perform Toffoli gates and always use higher level states to improve fidelities for quantum circuits.  We find that these methods improve expected fidelities with increases of 2x across circuit sizes using intermediate encoding, and increases of 3x for fully-encoded ququart compilation.
\end{abstract}

\renewcommand{\shortauthors}{A. Litteken, L. Siefert, J. Chadwick, N. Nottingham, T. Roy, Z. Li, D. Schuster, F. Chong, J. Baker}




\maketitle


\begin{figure}[H]
    \centering
    \scalebox{0.98}{
    \includegraphics[width=\linewidth]{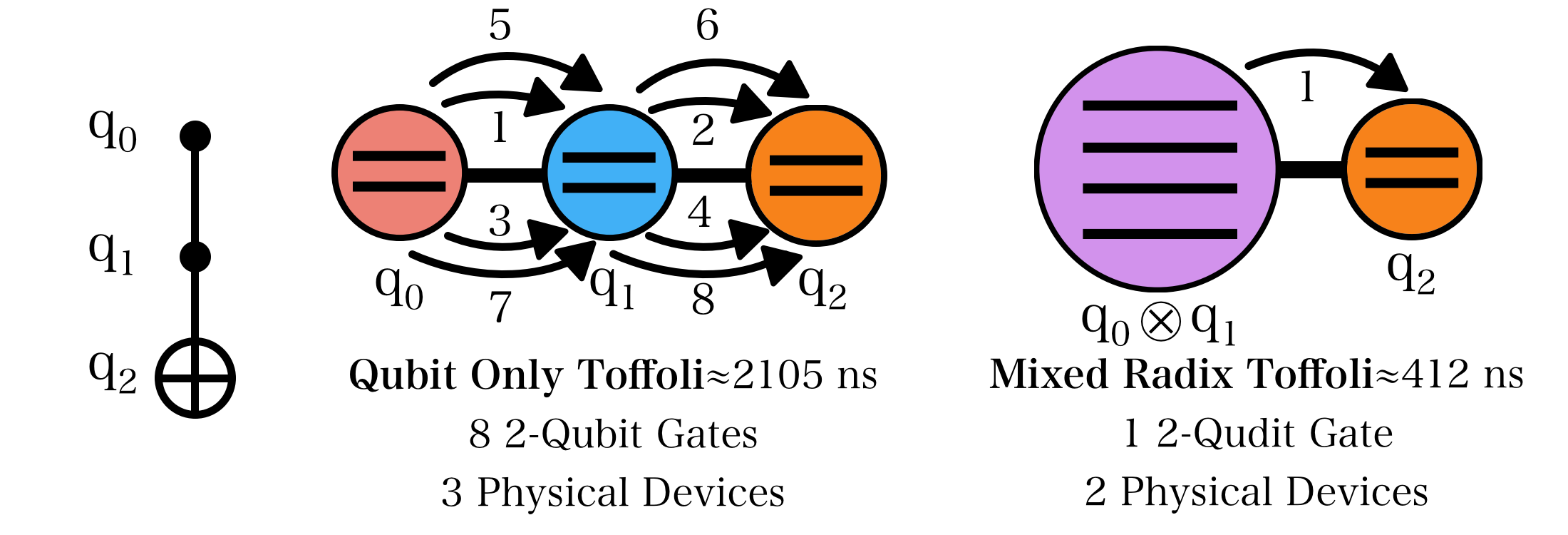}
    }
    \caption{A comparison of a Toffoli gate execution on a three-qubit-only system versus a Toffoli gate execution on a ququart and qubit in a mixed-radix system. In a qubit-only system, we must use a decomposition that uses eight two-qubit gates that can be reduced to one two-qudit gate that has a shorter duration.}
    \label{fig:qubit-vs-mixed-radix}
    \vspace{-0.5em}
\end{figure}

\section{Introduction}

Quantum systems are rapidly developing - stimulating the design and optimization of available hardware to maximize utilization of current resources for near-term quantum algorithms and the transition to quantum error correction \cite{gambetta_expanding_2022, chapman_scaling_2020}. Error-prone gates, sparse connectivity and low coherence times (the approximate computation time of any given device) are challenges currently facing quantum systems \cite{preskill_quantum_2018}. Even smaller quantum algorithms which fit on current hardware push devices to their limit. These limitations require improved optimization frameworks to make them useful in the near-term, prior to quantum error correction.

Success of quantum algorithms depends on how many error-prone gates are used and the total program duration. In most competitive quantum systems, e.g. superconducting systems, trapped ions, and neutral atoms, gates which act on many qubits simultaneously ($\ge 3$ operands) must be decomposed, increasing both gate counts and circuit depth. In this work, we focus primarily on superconducting systems, where limited connectivity between devices further exacerbates the decomposition problem.  Many circuits include gates, such as the Toffoli gate, to perform reversible arithmetic calculations; thus, three-qubit operations are common across implementations of quantum algorithms \cite{baker_efficient_2020, grover_fast_1996, cuccaro_new_2004, gokhale_quantum_2020}. Finding gate implementations without having to reduce them to more elementary gates saves valuable computational resources.

Currently, most quantum devices use qu\textit{bits}, which have two energy levels, used to represent the $\ket{0}$ and $\ket{1}$ state. Recently, there have been several explorations into using the higher energy levels such as $\ket{2}$ and $\ket{3}$ to reduce the number of gates required to perform computation. While there are many examples of exploiting this concept of qu\textit{dits}, such as using qutrits (3 logical states) to implement the multi-control Toffoli gate \cite{gokhale_asymptotic_2019, litteken_communication_2022}, implementing higher-radix adders \cite{taheri_monfared_quaternary_2019}, and other applications \cite{ivanov_time-efficient_2012}, these use cases are the result of hand optimization, making their general use limited.

Another proposed use of higher-radix states is to fully encode data from two qubits into one physical unit with four logical levels, called a ququart.  Previously, this strategy was avoided due to more error-prone operations \cite{chi_programmable_2022, wang_qudits_2020} and lower coherence times. 
Though coherence time is a limited resource, we can solve this problem by developing a set of operations which make better use of additional logical levels.

In this work, we observe that one ququart is equivalent to two qubits; thus the information of two qubit devices can instead occupy a single device which has access to four logical states \cite{baker_efficient_2020}. This has significant advantages on the relative connectivity of the qubit information: by performing this compression, we can access three qubits worth of information by interacting only two physical devices in a single operation rather than directly interacting three physical devices in a single operation. This type of \textit{mixed-radix} gate (four-level system interacting with an adjacent two-level system) is equivalent to performing a three-qubit gate. Similarly, we can consider two adjacent ququarts which allows us to perform interactions on up to four qubits worth of information by controlling only two physical devices; we call these \textit{full-ququart} gates. Our strategy could remove the need to perform expensive decompositions of three- or four-qubit gates, as visualized in Figure \ref{fig:qubit-vs-mixed-radix} potentially improving circuit fidelity of circuits containing multiqubit gates through the direct execution of three-qubit gates. We are primarily focused on common three-qubit interactions since they appear more commonly in real applications, unlike four-qubit gates.

\begin{figure}
    \centering
    \scalebox{0.85}{
    \includegraphics[width=\linewidth]{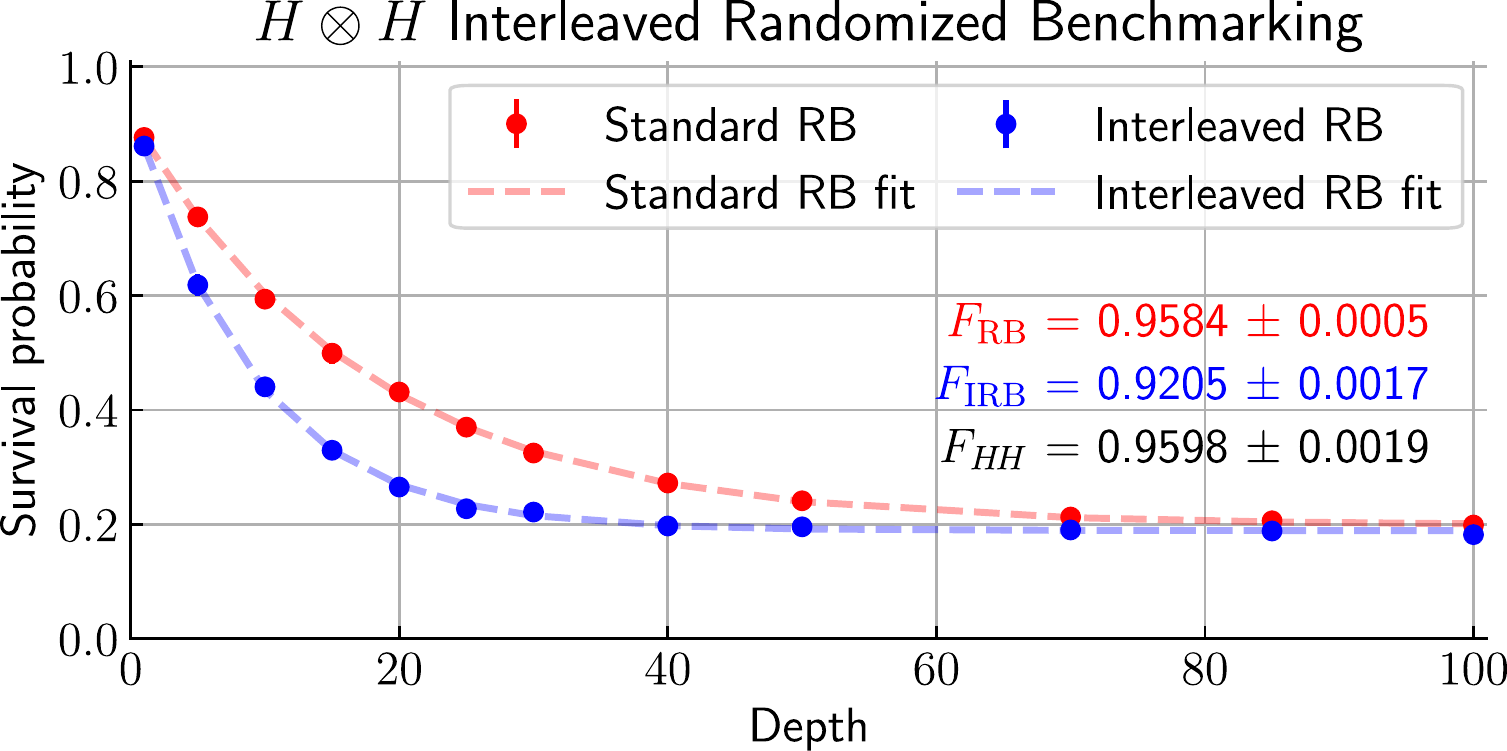}
    }
    \caption{Interleaved Randomized Benchmarking for an optimal control $H \otimes H$ pulse on a superconducting transmon ququart following our qubit encoding. We use two-qubit Clifford sequences of gate depth up to 100 and average each data point over 10 samples. Error bars show the standard deviation of the mean but they are smaller than the mean markers. Red: Standard two-qubit Randomized Benchmarking to estimate the average Clifford gate fidelity to be $F_\mathrm{RB} \approx 95.8\%$. Blue: Interleaving the $H \otimes H$ pulse between the RB Cliffords yields a combined per-operation fidelity of $F_\mathrm{IRB} \approx 92.1\%$, resulting in an $H \otimes H$ fidelity $F_{H \! H} \approx 96.0\%$.}
    \label{fig:HH_IRB}
    \vspace{-1.0em}
\end{figure}

We examine using ququarts to dynamically encode and decode gates to perform native three-qubit gates on ququarts on a simulated superconducting device in a compilation pipeline called the Quantum Waltz, a dance done in three-four time.  In particular, the major contributions are the following:
\begin{itemize}[leftmargin=*]
    \item A collection of mixed-radix and full-ququart gates that are logically equivalent to qubit-only gates, allowing for translation between qubit and mixed-radix operation.
    \item Demonstrating viability of ququart operations via pulses generated optimal control on hardware not previously designed for ququart pulses, Figure \ref{fig:HH_IRB}
    \item Identifying specific relationships between the controls and targets of three-qubit gates that allow for more efficient execution of mixed-radix and full-ququart gates with a compiler that choreographs three-qubit gates into particular configurations on ququarts for better performance and as a viable alternative to qubit-only strategies.
    \item Demonstrating, in simulation, how three-qubit gates on ququarts can achieve a 2x improvements in simulation in a mixed-radix environment and up to 3x fidelity improvements in a full ququart environment, as well as insights into the right situations to implement these gates.
\end{itemize}

\section{Background}

\subsection{Quantum Circuits}
Quantum computation focuses on the use and manipulation of the qubit states $\ket{0}$ and $\ket{1}$, which can exist in a superposition of these states as $\ket{\psi} = \alpha\ket{0} + \beta\ket{1}$ prior to measurement. $N$ qubits exist in a superposition of $2^N$ basis states given by bitstrings of length $N$. These states are manipulated through the use of quantum logic gates in quantum circuits.

In principle, gates can act on any number of qubits. We mainly focus on single-, two- and three-qubit gate.  Multi-qubit gates often use controls, meaning the state of another qubit only changes when the value of the other qubits is in a specific state. For three-qubit gates, this can mean there are multiple controls, like the Toffoli gate shown in Figure \ref{fig:qubit-vs-mixed-radix}.  Similarly, multiple qubits can be controlled by one qubit. For comprehensive review of quantum gates we refer to \cite{nielsen_quantum_2011}.  

\subsection{Higher Radix Computation}
Most abstractions of quantum computing are binary, focusing on the superposition of only two computational states. Many physical quantum technologies have access to higher energy levels which can be used to represent additional logical states as qu\textit{dits} which use the lowest $d-1$ energy states which are increasingly harder to control. In this work, we constrain ourselves to at most four logical states, a \textit{ququart}, which balances the potential computational benefit with its increasing error and time cost. In its naive use-case, additional levels have the same computational benefit as in classical - at most constant reductions in circuit depth and gate counts \cite{pavlidis_arithmetic_2021}.

Some work \cite{gokhale_asymptotic_2019, litteken_communication_2022} has demonstrated specific applications that take advantage of extra computational states to reduce space requirements and improve execution time.  These strategies are not generally applicable as it requires hand optimization for those circuits.  Other work \cite{baker_efficient_2020} attempted to generalize these improvements through compression, which stores multiple qubits worth of information in a smaller number of qudits. However, the usefulness of this strategy for general applications has not been explored and did not consider direct-to-pulse implementations of multi-qudit gates.

\begin{table*}[htbp]
    \centering
    \caption{Durations for one-qubit, two-qubit and $i$Toffoli gates synthesized in qubit-only, mixed-radix and full-ququart environments.}
    \renewcommand{\arraystretch}{1.2}
    \begin{tabu}{l r|l r|[2pt]l r|[2pt]l r|l r|[2pt]l r|l r}
        \multicolumn{4}{c|[2pt]}{\textbf{(a) Qudit (ns)}} & \multicolumn{2}{c|[2pt]}{\textbf{(b) Qubit Only (ns)}} & \multicolumn{4}{c|[2pt]}{\textbf{(c) Mixed-Radix (ns)}} & \multicolumn{4}{c}{\textbf{(d) Full-Ququart (ns)}}\\
        \hline
        
        U & 35 & U$^0$ & 87 & CX$_2$ & 251 & CX$^{0q}$ & 560 & CX$^{q0}$ & 880 & CX$^{00}$ & 544 & CX$^{01}$ & 544 \\

        U$^1$ & 66 & U$^{0,1}$ & 86 & CZ$_2$ & 236 & CX$^{1q}$ & 632 & CX$^{q1}$ & 812 & CX$^{10}$ & 700 & CX$^{11}$ & 700 \\

        CX$^0$ & 83 & CX$^1$ & 84 & CS$^\dagger_2$ & 126 & CZ$^{q0}$ & 384 & CZ$^{q1}$ & 404 & CZ$^{00}$ & 392 & CZ$^{01}$ & 488 \\
        
        SWAP$^{in}$ & 78 & & & SWAP$_2$ & 504 & SWAP$^{q0}$ & 680 & SWAP$^{q1}$ & 792 & CZ$^{11}$ & 776 & SWAP$^{00}$ & 916 \\

        &&&& $i$Toffoli$_3$ & 912 & ENC & 608 &&& SWAP$^{01}$ & 892 & SWAP$^{11}$ & 964 

    \end{tabu}
    \label{tab:two-qubit-gates}
    \vspace{-0.4em}
\end{table*}

\subsection{Quantum Optimal Control}
The state of qudits is manipulated through external hardware-specific control fields $f_k(t)$. We consider superconducting devices, so these control fields are analog microwave pulses. Given a target unitary operation $U$, quantum optimal control finds controls $f_k$ which realize $U$. Many optimal control algorithms and toolboxes have been developed \cite{khaneja_optimal_2005, sklarz_loading_2002, petersson_optimal_2021, gunther_quantum_2021}, and here we make use of the open-source software package Juqbox \cite{petersson_discrete_2020, petersson_optimal_2021}. We find control pulses of shortest duration which realize gates of interest up to competitive fidelity, $0.99$ for two-qudit gates and $0.999$ for single-qudit gates. Juqbox achieves this by minimizing the objective $J[f_k] = 1 - F[f_k] + L[f_k]$ where

\begin{equation}
    F[f_k] = \frac{1}{h^2} \abs{\Tr{U^\dagger_T[f_k] \, V}}^2
\end{equation}
quantifies the gate fidelity between target unitary $V$ and the applied transformation $U_T[f_k]$. Here $h$ is the Hilbert space dimension of the logical subspace (in our case $h=d$) and $T$ denotes the allotted gate time. This task is solved by repeatedly solving the Schrödinger equation and adjusting the control fields to minimize $J$. Higher energy levels are sometimes included in the simulation in order to accurately capture their effect on the state evolution and reduce errors from truncating high-dimensional systems. These guard states are not logical states, therefore populating them is penalized with a leakage term $L[f_k]$. Currently, Juqbox only allows pulse optimization for a fixed gate time $T$, therefore we minimize pulse durations by applying an iterative re-optimization technique \cite{seifert_time-efficient_2022}.  

\section{Compression and Gate Set}

\subsection{Information Compression}
Information \textit{compression} in the context of this work refers to the storage of many qubits worth of information which deviates slightly from the typical classical understanding. The goal of this compression is to reduce the total number of physical units required to realize a given quantum algorithm. 

Rather than designing algorithms that specifically use higher level states, we encode the data of two individual qubits into one four-level computational unit, called a ququart, given as $\ket{\psi}_4 = \alpha\ket{0} + \beta\ket{1} + \gamma\ket{2} + \delta\ket{3}$. This can be seen as equivalent to $\ket{\psi_1}_2 \otimes \ket{\psi_2}_2 = \alpha_1\alpha_2\ket{00} + \alpha_1\beta_2\ket{01} + \beta_1\alpha_2\ket{10} + \beta_1\beta_2\ket{11}$ by the following mapping:
\begin{align*}
       \ket{00} \rightarrow \ket{0} \ \  \ket{01} \rightarrow \ket{1} \ \ \ket{10} \rightarrow \ket{2} \ \ \ket{11} \rightarrow \ket{3}
\end{align*}
Therefore, $\alpha = \alpha_1\alpha_2$, $\beta = \alpha_1\beta_2$ etc.  This compression does not result in the loss of any information since the transformation from $\ket{\psi_1}_2 \otimes \ket{\psi_2}_2$ to $\ket{\psi}_4$ is unitary and therefore invertible.  This follows a modification of the scheme from \cite{baker_efficient_2020}.
 
This compression does not require a circuit to explicitly use the $\ket{2}$ and $\ket{3}$ states for the compiler to make use of ququarts like in \cite{gokhale_asymptotic_2019, ivanov_time-efficient_2012, wang_qudits_2020}.  We are able to adapt qubit-compiler pipelines to compile a circuit and encode qubits into a ququart and keep track of the original qubits without requiring changes in the original circuit.

\subsection{Qubit Gates on Ququarts}
\label{sec:qubit-gates-on-ququarts}
Past work has studied higher level systems by generalizing operations on a qubit circuit. For example, the X gate, is generalized to a $+1 \mod d$ instead, where $d$ is the dimension of the qudit. Multi-qubit gates generalize similarly; for example a CNOT can be viewed as a $\ket{1}$-controlled $+1 \mod 2$ gate and therefore in general we can consider $\ket{c}$-controlled $+m \mod d$ gates, $0 \leq c, m \leq d - 1$ \cite{luo_universal_2014}. 

While possible to use this generalized gate set to perform computation, it is not concise. For example, to perform a CNOT between the second encoded qubits encoded in different ququarts we would need to apply two $\ket{1}$-controlled $+1$ gates and two $\ket{3}$-controlled $+1$ gates. We could instead generate and calibrate a more expressive gate set that directly performs this operation.

We develop a gate set which performs qubit operations directly on ququarts.  For a single-qubit gate $U$ acting on two encoded qubits in the state $\ket{q_0q_1}$, we use the unitary $U^0 = U \otimes \mathbbm{1}$ to act on qubit $q_0$, $U^1 = \mathbbm{1} \otimes U$ to act on qubit $q_1$, and $U^{0,1} = U \otimes U$ to act on both qubits simultaneously.

For two-qubit gates, there are several important classes of operations. The first is the interaction between the two compressed qubits which we call an \textit{internal} operation. For example, a $\text{CX}^0$ is a CNOT controlled on the second qubit targeting the first; this is equivalent to the single ququart gate which swaps the states $\ket{1}$ and $\ket{3}$. $\text{CX}^1$ controls on the first and targets the second encoded qubit. A SWAP operation exchanges the order of the encoding, i.e. SWAP$\ket{q_1q_2} = \ket{q_2q_1}$. The second are gates which act on qudits in different, but adjacent, physical locations. These \textit{partial} gates interact a non-encoded qubit and a qubit in an encoded pair in adjacent locations; all gates of this type we call \textit{mixed-radix} gates. For these gates, order matters, i.e. the gate behaves differently depending on which qubit is the target.
  
The four CX gates are $\{\text{CX}^{q0}, \text{CX}^{q1}, \text{CX}^{0q}, \text{CX}^{1q}\}$ where the first index indicates the control and the second the target object, and $q$ is the qubit.  We also define two mixed-radix SWAPs $\{\text{SWAP}^{q0}, \\ \text{SWAP}^{q1}\}$ which are the same regardless of direction.  The \textit{ full-ququart} gates follow from the mixed-radix gates defining the four CX gates: $\{\text{CX}^{00}, \text{CX}^{01}, \text{CX}^{10}, \text{CX}^{11}\}$ and three SWAPs: $\{\text{SWAP}^{00}, \\ \text{SWAP}^{01}, \text{SWAP}^{11}\}$.

\subsection{Generating Pulses}\label{sec:pulse-generation}
Using quantum optimal control we directly synthesize each of the gates in our new mixed-radix and full-ququart gate set and baseline comparisons. We use a realistic superconducting device Hamiltonian inspired by IBM hardware \cite{sheldon_procedure_2016}.

We consider up to three weakly coupled, anharmonic transmons \cite{koch_charge-insensitive_2007}:
\begin{align}
    &H(t) = ~\sum_{k=1}^3 \qty[\omega_k a_k^\dagger a_k + \frac{\xi_k}{2} a_k^\dagger a_k^\dagger a_k a_k] \\
    &+ \sum_{k=1}^3\sum_{l>k}J_{kl} (a_1^\dagger a_2 + a_2^\dagger a_1) \notag 
    + \sum_{k=1}^3 f_k(t) (a_k + a_k^\dagger).
    \label{eq:ham_rot}
\end{align}

The static terms describe the individual qudits and their pairwise couplings, while the last term captures the effect of driving the system through external control fields $f_k(t)$. The transmons are designed with $\ket{0}$-$\ket{1}$ transition frequencies $\omega_1/2\pi = 4.914 \,\mathrm{GHz}$, $\omega_2/2\pi = 5.114 \,\mathrm{GHz}$, and $\omega_3/2\pi = 5.214 \,\mathrm{GHz}$, and with equal anharmonicities $\xi_k/2\pi = -330 \,\mathrm{MHz}$. We consider linear connectivity with static couplings given by $J_{12}/2\pi = J_{23}/2\pi = 3.8\,\mathrm{MHz}$. The drive power is limited to $f_\mathrm{max} = 45\,\mathrm{MHz}$ to avoid substantial leakage into higher energy states, and we restrict ourselves to the $k=1$ subspace when synthesizing single-qudit gates. 

A full list of the gates synthesized and the minimal found duration of these gates can be found in Table \ref{tab:two-qubit-gates}. We reiterate the importance of short gate times - quantum systems are subject to a variety of both coherent and incoherent errors. By minimizing the total execution time of any given gate we reduce the circuit duration, reducing the effects of incoherent noise.

The closed system considered does not account for the full dynamics of a real quantum device. We have not specifically optimized these pulses under a more detailed model due to the increased computational cost of these optimizations, especially for the large Hilbert spaces involved in two-qudit operations.

In Section \ref{sec:experiment}, we use similar optimal control techniques to implement a single-ququart operation on an experimental device, showing that our methods and assumptions are realistic given a well-characterized machine.

\subsection{Properties of Qubit Gates on Ququarts}
\begin{figure}
    \centering
    \includegraphics[width=0.7\linewidth]{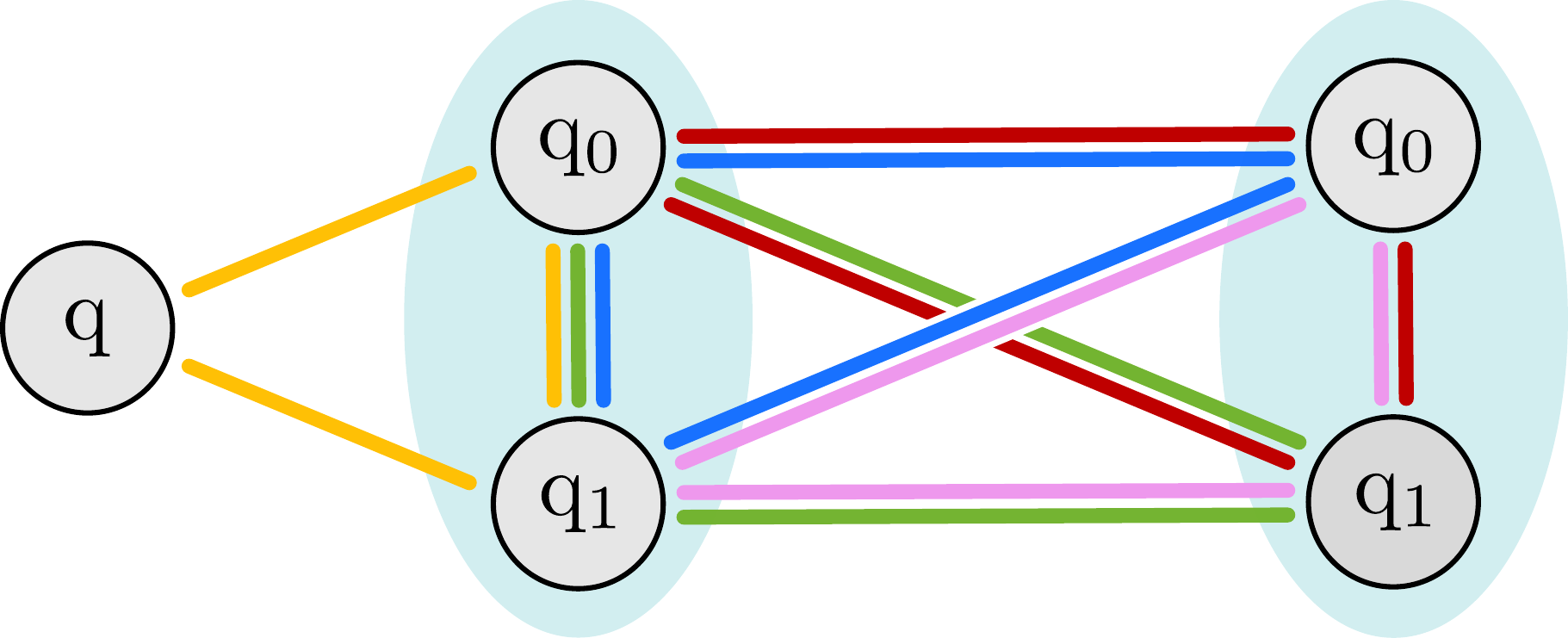}
    \caption{Visualization of connectivity advantages in qubit-ququart systems. Encoding qubits in ququarts (light blue) enables triangle connectivity between triples of qubits, where two of which are encoded in the same ququart and one appears either in a bare qubit or encoded in a neighboring ququart.}
    \label{fig:mixed-radix-and-full-encoded}
\end{figure}

Our gate set and mixed-radix architecture provides real advantages over typical qubit-only versions. Within each ququart, we have a pair of encoded qubits between which gates are 5x faster and 10x higher fidelity than qubit-only schemes. By using a single computation device, the total amount of control hardware required is reduced (at most by half).  Additionally, we have much higher connectivity between qubits once they are encoded in ququarts. In a ququart-qubit pair, there are three computational qubits directly connected to one another. Between two ququarts, there are four fully connected computational qubits. This is higher relative connectivity compared to industry standards for superconducting: lines, grids, and heavy hex architectures. Improved connectivity reduces expensive qubit movement operations.  These increased connections are demonstrated in Figure \ref{fig:mixed-radix-and-full-encoded}.

Compression is not without its downsides. In Table \ref{tab:two-qubit-gates}, we see mixed-radix and ququart gates take much longer than qubit based gates. Pulses must be more carefully designed, and leakage between states is more prominent, resulting in the longer gates times. Each increasing energy level has a shorter coherence time scaling with $1/k$ where $k$ is the energy level. Shorter decoherence, combined with longer gate times, means using mixed-radix and ququart based gates is a delicate balancing act between increasing fidelity due to gate execution while not increasing error due to decoherence.

\subsection{Experimental Demonstration of Single-Ququart Control}\label{sec:experiment}

Driven by advantages found in theoretical studies \cite{pavlidis_arithmetic_2021}, experimental researchers have explored the implementation of these higher-dimensional systems, leading to realizations of qutrit devices which manipulate the third energy level \cite{galda_implementing_2021, hill_realization_2021, roy_realization_2022, goss_high-fidelity_2022, morvan_qutrit_2021, wu_high-fidelity_2020}. These works show that including higher levels is possible although challenging due to higher susceptibility to noise and lower coherence times.

Motivated by the findings for ququart-specific applications we have been studying control of four energy levels in experiment on a physical device. We extend the capabilities of one qudit of the superconducting transmon device presented in \cite{li_autonomous_2023, li_hardware_2023, roy_realization_2022} to include the fourth state. We implement two-qubit Randomized Benchmarking (RB) \cite{magesan_scalable_2011} on this single ququart following our encoding scheme. RB is a common method to characterize the average Clifford gate fidelity. This is achieved by executing Clifford circuits of varying depth, which perform the identity operation in the ideal case, and measuring the probabilities of the system returning to the ground state (survival probabilities). The fidelity can be extracted from exponential regression. The RB circuits are generated using Qiskit \cite{anis_qiskit_2021}.

We additionally implement Interleaved Randomized Benchmarking (IRB) \cite{magesan_efficient_2012} to specifically find the fidelity of the single-ququart gate $H \otimes H$, which performs a Hadamard gate on each encoded qubit in parallel, used by the compiler below. The gate control pulse is designed using similar optimal control methods as discussed in Section \ref{sec:pulse-generation} adapted to this experimental device.

Results from this work are shown in Fig. \ref{fig:HH_IRB}. We find an average Clifford gate fidelity of $F_\mathrm{RB} \approx 95.8\%$ from normal RB while interleaving with the $H \otimes H$ gate yields $F_\mathrm{IRB} \approx 92.1\%$ fidelity per operation. From that the specific gate fidelity, $F_{H \! H} \approx 96.0\%$, can be extracted. This first study shows that ququarts can be realized in experiment and optimal control yields high-quality pulses to manipulate their state. At the time of this writing we are not aware of any comparable demonstration. We are convinced that the fidelities can be improved with more carefully engineered ququart devices and more sophisticated pulse design methods.

\section{The Quantum Waltz: Three Qubit Gates on Ququarts}

Three-qubit gates are widely used in arithmetic operations, such as the Cuccaro adder \cite{cuccaro_new_2004} and multi-controlled-CNOT \cite{barenco_elementary_1995}, as smaller pieces in larger quantum algorithms such as \cite{grover_fast_1996}.  While QAOA and VQE see more use in the current quantum algorithm space, some QAOA based algorithms still use three-qubit gates \cite{hadfield_quantum_2019}.  Additionally, some error correction schemes make heavy use of three-qubit gates \cite{yoder_universal_2017}.  Current hardware platforms typically decompose these gates. Using higher radix we can reduce gate times mitigating the issue of reduced coherence times of higher-energy levels enabling more efficient execution of quantum circuits containing these gates.

\subsection{Connectivity Advantage}

The set of two qubit gates laid out in Section \ref{sec:qubit-gates-on-ququarts} are enough to universally perform general qubit computation on ququarts \cite{luo_universal_2014, nielsen_quantum_2011}, but simply compiling to two-qubit gates would not take advantage of the flexibility of this abstraction. When we encode two qubits in a ququart, we virtually increase the connectivity between qubits, see Figure \ref{fig:mixed-radix-and-full-encoded}.  Each of the encoded qubits in a ququart is connected to an adjacent qubit, or both of the encoded qubits in an adjacent ququart. As highlighted by each of the different colors this creates many triangle subgraphs between encoded qubits . Triangle subgraphs are uncommon in current hardware due to the increased probability of crosstalk \cite{mundada_suppression_2019, ding_systematic_2020}. But, triangle-based interactions are common in many different circuits that use three-qubit gates.  Here, we increase the number of virtual connections without increasing number of physical connections to create four interactions between encoded qubits.

\begin{figure}
    \centering
    \includegraphics[width=0.85\linewidth]{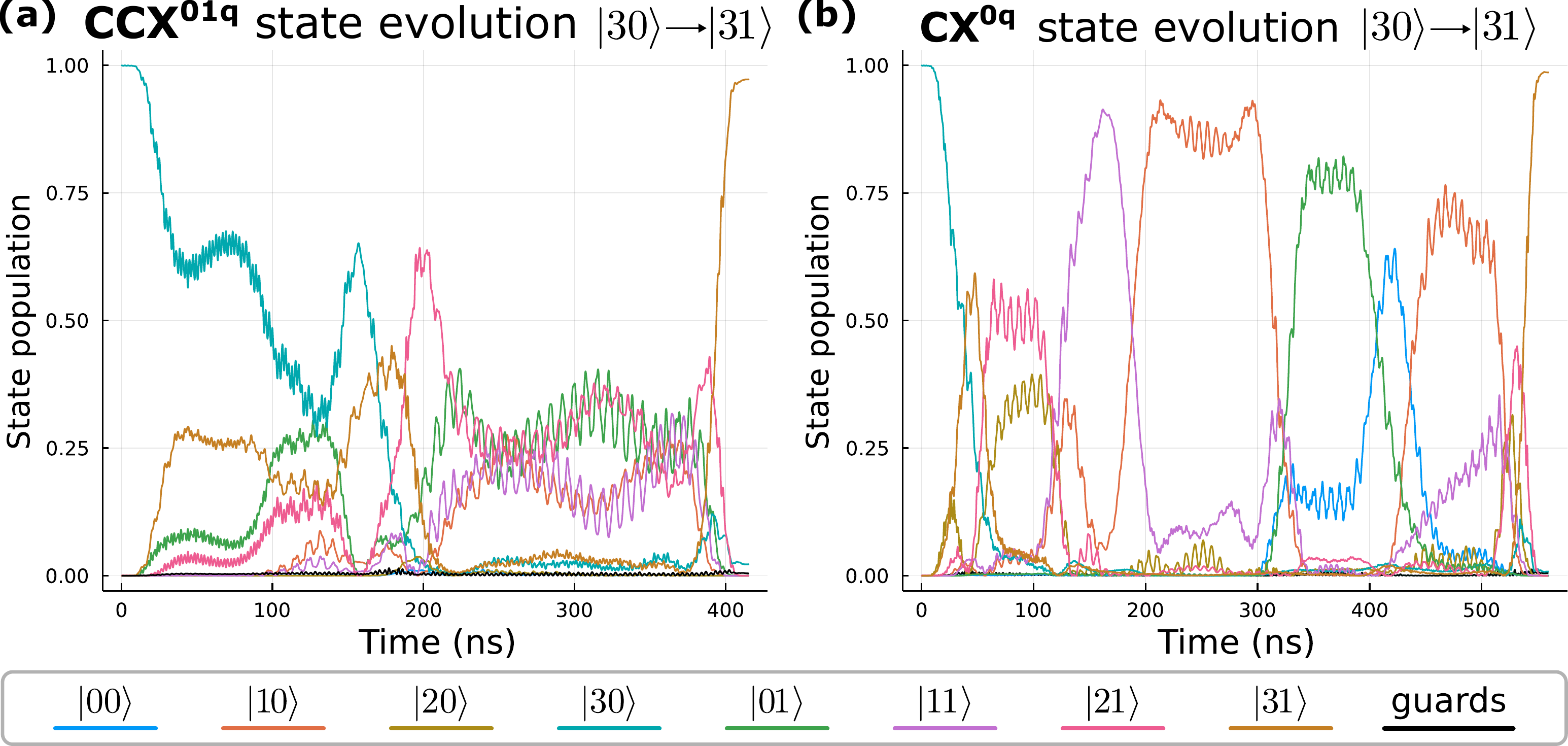}
    \caption{Visualization comparing the evolution of a $\ket{3}$-controlled $X$ gate in a mixed-radix environment for a CCX gate in (a) and a CX gate in (b).}
    \label{fig:state-evolutions}
\end{figure}

It is not fundamentally harder to interact three or four qubits worth of information than two qubits worth with a single operation on ququarts. These gates are equivalent to either mixed-radix or full-ququart gates. For example, if we have a fully encoded ququart next to a bare qubit and perform a Toffoli gate targeting the qubit, it is equivalent to a $\ket{3}$-controlled X  gate on the qubit. This is computationally simpler than the several $\ket{1}$- and $\ket{3}$-controlled X required in the decomposition and can be seen in the state evolutions in Figure \ref{fig:state-evolutions}. This gate implementation gives superconducting qubits more natural access to the native multi-qubit gates,  avoids decompositions that add extra gates and performs three-qubit interactions between two physical quantum devices, reducing the complexity of implementing such a three-qubit pulse across three devices and two couplers. Used in conjunction with the previously generated one- and two-qubit gates, we can more efficiently perform circuits that include three-qubit gates.

\subsection{Generated Pulses}

\begin{table}[htbp]
    \centering
    \caption{Mixed-Radix and Full-Ququart Three-Qubit Gate Durations}
    \renewcommand{\arraystretch}{1.2}
    \begin{tabu}{l r|[2pt]l r|l r}
        \multicolumn{2}{c|[2pt]}{\textbf{(a) Mixed-Radix (ns)}} & \multicolumn{4}{c}{\textbf{(b) Full-Ququart (ns)}} \\
        \hline
        CCX$^{q01}$ & 619 & CCX$^{01,0}$ & 536 & CCX$^{01,1}$ & 552 \\
        CCX$^{1q0}$ & 697 & CCX$^{0,01}$ & 785 & CCX$^{0,10}$ & 785 \\
        CCX$^{01q}$ & 412 & CCX$^{1,10}$ & 785 & CCX$^{1,01}$ & 680 \\
        \hline
        CCZ$^{01q}$ & 264 & CCZ$^{01,0}$ & 232 & CCZ$^{01,1}$ & 310 \\
        \hline
        CSWAP$^{01q}$ & 684 & CSWAP$^{01,0}$ & 680 & CSWAP$^{01,1}$ & 744 \\
        CSWAP$^{10q}$ & 762 & CSWAP$^{10,0}$ & 758 & CSWAP$^{10,1}$ & 822 \\
        CSWAP$^{q01}$ & 444 & CSWAP$^{0,01}$ & 510 &  CSWAP$^{1,01}$ & 432
    \end{tabu}    
    \label{tab:three-qubit-gate-times}
\end{table}

\subsubsection{Multi-control Gates}
Native three-qubit gates on two physical units have the potential to offer a significant improvement in gate fidelity and execution time.  In Table \ref{tab:three-qubit-gate-times}, we show pulse durations of the three-qubit Toffoli gate in several mixed-radix and full-ququart configurations. These gates were synthesized using the same fidelity targets and pulse generation techniques as the two-qubit gates, a higher fidelity than if decomposed with many gates of the same target fidelity. After synthesizing the different configurations of Toffoli gates, we find that there is a substantial difference in the gate duration depending on which qubits are controls and which is the target.

\begin{figure}
    \centering
    \scalebox{0.85}{
    \includegraphics[width=\linewidth]{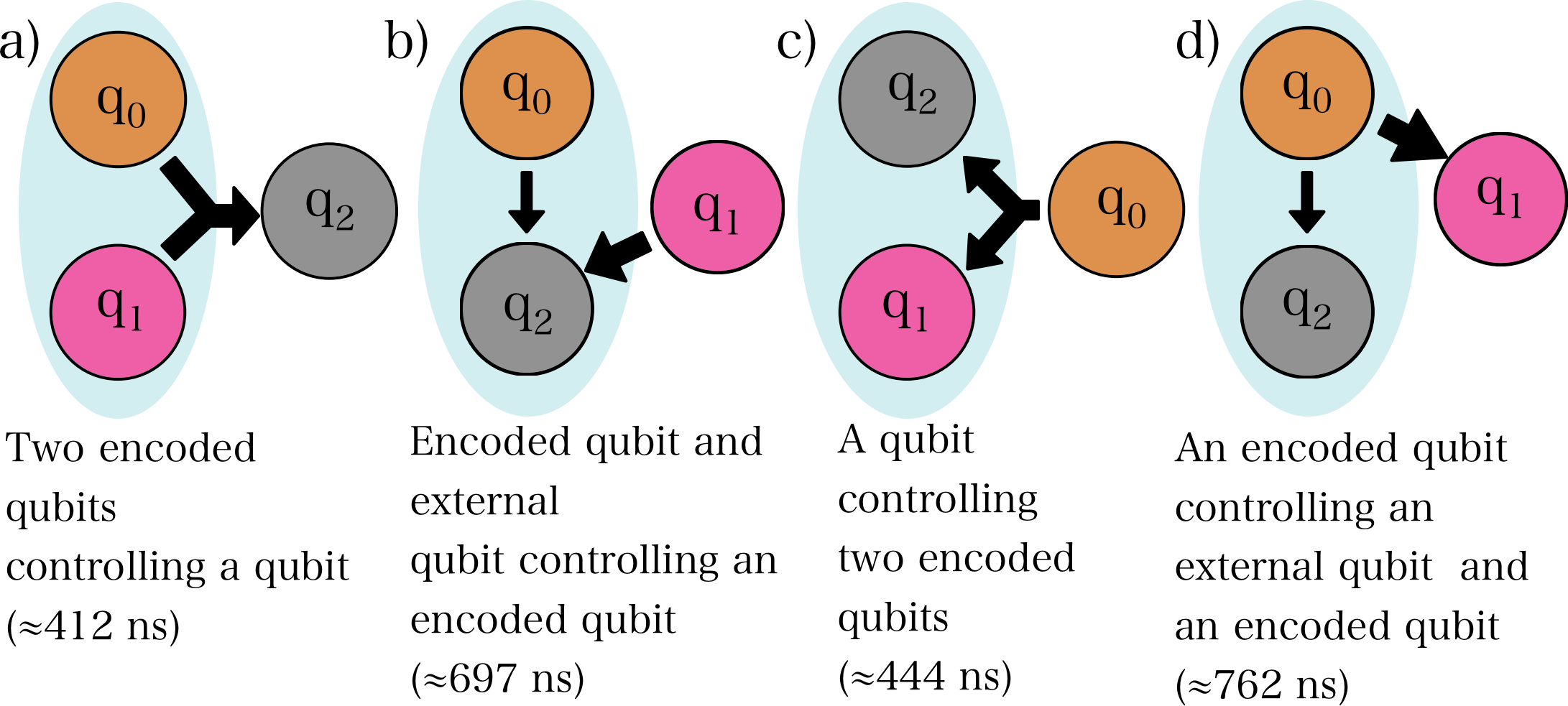}
    }
    \caption{Examples of mixed-radix two-control and two-target gates. a) A configuration where both controls are encoded in the ququart and the target is mapped to a qubit. b) A configuration where the controls are split across the qubit and the ququart and the target is encoded in the ququart. c) A configuration where both targets are encoded in the ququart and the control is mapped to the qubit. d) A configuration where the targets are split across the qubit and the ququart and the control is encoded in the ququart.}
    \label{fig:three-qubit-configs}
    \vspace{-0.5em}
\end{figure}

Consider the mixed-radix example where both control qubits are encoded in the same ququart, and the target qubit is in the bare qubit, or the $\text{CCX}^{01q}$ gate, seen in Figure \ref{fig:three-qubit-configs}a.  This configuration is about two-thirds the time of the $\text{CCX}^{0q1}$, seen in Figure \ref{fig:three-qubit-configs}b, where the control qubits are split across the bare qubit and the ququart.  The reason for this difference is twofold.  The first follows from the two-qubit only gates. Gates which use the ququart as a control and the qubit as a target are generally faster, the pulse only induces population changes between the $\ket{0}$ and $\ket{1}$ state of the qubit, rather than between $\ket{0}$ and $\ket{1}$, and $\ket{2}$ and $\ket{3}$ of the ququart.  The second is that the entire ququart acts as the control, only changing the state of the bare qubit if the ququart is in the $\ket{3}$ state.  In the split-control case, the ququart must control on both the $\ket{2}$ and $\ket{3}$ state.

The same concept of separation of controls and targets follows for the full-ququart Toffoli gates as well.  Regardless of whether the target qubit is in the first or second encoding of the ququart, it is substantially faster to keep the controls encoded in the \textit{same} ququart with the target encoded in a separate ququart. 

\subsubsection{Target-Independent Gates}
Separating the controls and targets into different devices yields more efficient gate execution; however, compiling circuits to conform to this configuration is unnecessarily constraining. Instead, we consider a situation where all multi-qudit gates are \textit{target-independent} and only affect the global state when all three qubits are in $\ket{1}$. For example, the Toffoli gate, or CCX, is locally equivalent to CCZ which is target-independent, as seen in Figure \ref{fig:toffoli_decomp}c. 

When pulses are synthesized, CCZ is much more efficient as seen in Table \ref{tab:three-qubit-gate-times}, remarkably on par with the speed of the qubit only gates. In addition, we only need to define three configurations: $\text{CCZ}^{q,01}, \text{CCZ}^{0,01}, \text{CCZ}^{1,01}$, as opposed to the nine possible CCX configurations, reducing computational overhead. We postulate the short duration of these gates is because CCZ only changes the phase of the entire three-qubit state rather than the population. This makes the CCZ a valuable tool when compiling three-qubit gates.

\subsubsection{Multi-target Gates}
\label{sec:multitarget}
We also consider gates that use one control qubit to affect the state of some number of other qubits, for example the CSWAP. With our methods we synthesize gates to the same fidelity targets as before and show their times in Table \ref{tab:three-qubit-gate-times}. We find benefits when separating the control qubit from the target qubits as depicted in Figure \ref{fig:three-qubit-configs}c versus Figure \ref{fig:three-qubit-configs}d.  When both targets are encoded the same ququart, we limit the state changes to be between $\ket{1}$ and $\ket{2}$ in that ququart.
\section{Compilation Strategies}

\subsection{Using Three-Qubit Gates}
In our qubits-on-ququarts compilation strategy, we expand the physical connectivity graph between the ququarts on a given architecture and treat each ququart as two connected qubits.  Each qubit in the expanded ququart is fully connected to the qubits in the neighboring ququarts as shown in Figure \ref{fig:mixed-radix-and-full-encoded}.  We call this new graph the interaction graph; it maintains a mapping of where circuit qubits are mapped to on this graph. When one or fewer of the qubits in the expanded ququart is mapped to, the entire ququart is in a qubit state. Otherwise, it is considered to be in the ququart state.

To execute three-qubit gates, circuit qubits must be routed into a connected subgraph of the interaction graph, e.g. for CCZ$(q_0, q_1, q_2)$ requires $q_0\sim q_1$ and $q_1\sim q_2$ but it is not guaranteed that $q_0\sim q_2$, where $\sim$ defines adjacency. We develop a compiler optimization which appropriately performs routing and gate selection based on this adjacency and use of higher dimension. While we are able to perform any configuration of three qubit gates directly in mixed-radix or full-ququarts scenarios, we take care to use best configurations to minimize time in the less stable $\ket{2}$ or $\ket{3}$ states.

\subsubsection{Qubit-Only}
\begin{figure}
    \centering
    \scalebox{0.9}{
        \input{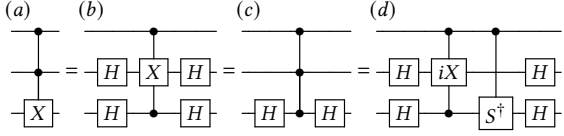}
     }
     \caption{Different decompositions for the Toffoli Gate. a) is the base Toffoli circuit. b) is Toffoli circuit with a swapped second control and target from the original.  By surrounding the control and the target with Hadamards, we perform the same operation. c) The Toffoli gate constructed from a CCZ gate which can be used as a Toffoli by surrounding the target with Hadamard gates. (d) The Toffoli gate constructed from an iToffoli gate, which requires an controlled $S^\dagger$ gate in addition to Hadamard gates.}
    \label{fig:toffoli_decomp}
    \vspace{-0.5em}
\end{figure}

In a qubit-only regime we can use a decomposition into eight CX operations \cite{shende_cnot-cost_2008}.  This decomposition has the flexibility of being target-independent from a compilation standpoint.
This is an expensive compilation, requiring eight two-qubit gates and 14 one-qubit gates.  But, it does not use the less stable $\ket{2}$ and $\ket{3}$ state.  Alternatively, we can use a directly-optimized three-qubit pulse sequence. QOC software failed to find a solution for a direct CCZ operation, so we synthesize a pulse implementing the $i$Toffoli gate using a three-qubit version of our quantum optimal control software that only uses the first two levels of the qubits and use the decomposition shown in Figure \ref{fig:toffoli_decomp}d inspired by \cite{kim_high-fidelity_2022} to execute a complete Toffoli gate.

\subsubsection{Intermediate Mixed-Radix}
We also permit \textit{temporary} use of the higher energy levels to perform an operation.  By performing an encoding gate (ENC) followed by the three-qubit gate and a final decode (ENC$^\dagger$) operation, we get temporary access to full connectivity to perform fast three-qubit gates.

The compiler should opt to encode qubits of similar \textit{type}, i.e. both controls together or both targets together. Let $U(q_0, q_1, q_2)$ be the operation with $q_0, q_1$ the same (either both controls or both targets). In some cases, encoding is simple because the routing strategy (prior work) results in $q_0\sim q_1$ as in  \ref{fig:three-qubit-configs}(a). However, it may fail to do this by default and we may have $q_0\sim q_2$ and $q_1\sim q_2$ as in \ref{fig:three-qubit-configs}(b). 

We have three options to compile to a favorable configuration.  First, we could enforce the ideal relationship through additional gates by adding an additional SWAP($q_0$, $q_2$). Second, in the special case where $U = X$ Toffoli we can change which pair is the same type with Hadamard gates as in Figure \ref{fig:toffoli_decomp}b to use the most efficient implementation; we call this \textit{re-targeting}. Third, if $U$ permits, we transform $U$ into $U'$ so that $q_0, q_1, q_2$ are all the same type; for example we transform CCX to CCZ so each operand is a ``control,'' Figure \ref{fig:toffoli_decomp}c. While the additional re-targeting or transformation gates add both error and duration, they enable the shortest duration version of $U$ to be used for an overall net increase in fidelity. We consider the special cases of $U \in \{CCV | V \in SU(2)\}$, i.e the set of locally equivalent gates to $CCX$. We leave the generalized case to future work in circuit synthesis.

\subsubsection{Full Ququart}
Mixed-radix three-qubit gate strategies apply for full-ququart compilation as well. However, the router by default, described below, does not distinguish control or target. When executing three-qubit gates, we ensure only qubits of the same type are encoded if it does not require an extra swap operation.

\subsection{Mapping and Routing}
Our compilation for encoded qubits on ququart architectures is similar to previous compilation strategies for qubits as seen in many previous works \cite{cowtan_qubit_2019, murali_noise-adaptive_2019, duckering_orchestrated_2021} and adapts them to three-qubit gates on ququart architectures.  However, unlike these prior works, we take into account the varying fidelities and durations of internal ququart versus mixed-radix versus full-ququart inter-ququart gates, similar to \cite{litteken_communication_2022}.

The first step is to decompose the operations in the circuit to native gates supported by the device. Our compiler handles the native execution of three-qubit gates, we decompose to the CX, CCX, CCZ or CSWAP along with a parameterized single-qubit rotation gate.

Qubits are mapped onto the interaction graph with the goal of maximizing locality.  We assign a weight between each pair of qubits in the original circuit according to: $w(i, j) = \sum_{t \in C} o(i, j, t)/{t}$, where the sum is over each time step $t$ in the circuit $C$ and $o(i, j, t) = 1$ if qubits $i, j$ interact in time step $t$ and $0$ otherwise. This weight includes lookahead functionality by weighting future interactions (larger $t$) smaller. 
The first qubit is mapped according to which has greatest total weight to all other qubits: \\ $\underset{i}{\text{argmax}} W(i) = \sum_{j \in Q_c / \{i\}} w(i, j)$.  This qubit is placed in the first encoded location of the center-most qudit on the connection graph.  For each other qubit, we choose the circuit qubit that has the greatest $W$ with respect to the placed qubits.  For each adjacent qubit, $n$, to the placed qubits, we compute $\sum_{j \in Q_P} w(i, j) d(n, \varphi(j))$ where $\varphi$ is the mapping of circuit qubits to physical qubits, and $d$ is a specialized fidelity function between the qubits estimating the possibility of error along the communication path.  We then map the qubit to the minimizing location.

When routing, we track the circuit qubits on the interaction graph and use SWAP gates until the interacting qubits are adjacent. We attempt to disrupt advantageous qubit layouts as little as possible by using adaptive weights that change as operations are scheduled based on \cite{baker_time-sliced_2020}.  This strategy attempts to keep qubits interacting in the near future close to one another where the disruption of each potential SWAP between circuit qubits $i, j$ is calculated by $D(i, j) = \sum_{k \in Q_c} w(i, k) (d(\varphi(i), \varphi(k)) - d(\varphi(j), \varphi(k))) +  w(j, k) (d(\varphi(j), \varphi(k)) - d(\varphi(i), \varphi(k)))$. However, rather than using simple distances, we use the same specialized distance metric incorporating the previous function $d$.  We choose the SWAP candidate that minimizes this value while always moving the qubit closer to the other qubits it needs to interact with.  To generalize to three-qubit based routing we modify the cost function to $C(i) = \sum_{j \in Q_o / i}D(i, \varphi^{-1}(n)) (d(\varphi(i), \varphi(j)) - d(n, \varphi(j))$  where $Q_o$ is now a set of all operands.

It would be reasonably simple to extend this compiler design to accomodate k-qubits on n-d-level-qudits, where we pack each qudit with $log_2(d)$ qubits, and ensure that there is no way to move any one qubit closer to another in a fully connected set of qubits. However, we only explore three-qubit gates on a maximum of two, four-level devices, or three, two-level devices in this work.  This is for design and practical reasons.  From a design point of view, our gate set and compiler are intended to be used after a circuit has been translated into qubit-based gates. Compiling natively to higher-radix qudit operations would require a much larger set of basis gates than the qubit-based set we use here.  Additionally, there is not a standard set of four-or-more qubit gates that are typically used in circuits, meaning there would have to be some arbitrary decomposition to four-qubit gates, rather than three qubits.  Choosing a basis gate set is a time intensive process and has to be done selectively \cite{gokhale_optimized_2020}. We therefore expand the normal one and two-qubit framework used in many compilers to include the most commonly used three-qubit gates as it is the most common multi-qubit gate.

It should be noted that all translation to higher-radix devices occurs during this compilation step.  The general programmer still writes a program in terms of qubits.  The compiler translates the program into the correct sequence of qubit-on-ququart operations to perform the same computation.  In the case of full-ququart operation, the measured state would be decoded according to the compression strategy.
\section{Evaluation}

\subsection{Circuits}

We examine five three-qubit based circuits that can be parameterized by number of qubits with different constructions. The first is the Generalized Toffoli (CNU) circuit \cite{baker_decomposing_2019}, which flips the state of a target qubit if all the controls are one. This circuit uses exclusively Toffoli gate based decomposition and is highly parallel. The Cuccaro Adder \cite{cuccaro_new_2004} is nearly entirely serialized using $2n+2$ qubits with a mix of three-, two- and single-qubit gates to add two $n$-bit numbers. Third is a QRAM circuit which uses primarily CSWAP gates to retrieve data from or move data into a set of qubits \cite{gokhale_quantum_2020}.  The fourth is a Select circuit, which is a preparation mechanism used in Quantum Phase Estimation (QPE) \cite{low_hamiltonian_2019}.  It performs a particular Pauli operation on $n$ qubits for each potential $2^m$ states of $m$ index qubits \cite{babbush_encoding_2018}.  For our case, the choice of Pauli string does not affect compilation.  To keep the fidelity of circuit simulation within comparable bounds, we only select on two random values rather than all of the potential $2^m$ values the index qubits could be in.  The fifth is a purely synthetic circuit to study relative strength of our architecture on potential distributions of CX versus CCX gates.

\subsection{Baselines, Hardware Topology and Error}
We compare against two strategies.  The first is a compilation that routes the circuit with three-qubit gates, before decomposing them to one- and two-qubit gates only.  This is in line with current practices for most compilation pipelines.  The second baseline does not decompose to these smaller gates.  Instead, the $i$Toffoli-based decomposition directly on qubits similar to \cite{kim_high-fidelity_2022}.  This is a more challenging gate to synthesize as discussed previously. For simulation, this gate has a 99\% fidelity and with 912 ns duration determined via the same quantum optimal control strategies as the mixed-radix and full-ququart gates.  Additionally, we use the Hadamard-based retargeting technique to ensure that we are applying the Toffoli gate to the correct qubit without an extra SWAP.  This allows us to always use the demonstrated $i$Toffoli gate where the target qubit is the center of three connected qubits.

We consider the same underlying hardware topology for each comparison point - a 2D mesh. This type of grid architecture has relative density on the upper end of realized superconducting connectivity graphs, reflective of Google's Sycamore chip \cite{arute_quantum_2019} and more dense than IBM's heavy-hex \cite{gambetta_expanding_2022}. We consider a  grid design with dimensions $\lceil\sqrt{n} \rceil \times \frac{n}{\lceil\sqrt{n}\rceil}$ with nearest neighbor connectivity.

We use a realistic T1 time from an IBM device of $163.45 \mu s$ \cite{ibm_ibm_nodate}.  Higher energy levels decohere more quickly.  In theory, each state decays at a rate of $o(1/k)$ where $k$ is the energy level as discussed in \cite{younis_berkeley_2021}. We therefore use $81.73 \mu s$ and $54.15 \mu s$ as the T1 times for the $\ket{2}$ and $\ket{3}$ states.  As any transmon technically has access to these higher energy states, we do not expect that a device designed to access these higher-energy states will reduce the base T1 time. 

\subsection{Circuit Estimation}
We use two metrics to estimate the fidelity of a circuit without simulation to extrapolate how compiled circuits may perform by comparing simulation to estimation. The first is the product of all of the gate success rates in the circuit, called the gate expected probability of success (gate EPS).  Since there are multiple classes of multi-qubit gates, some of which have higher fidelity than others, we use the product of these success rates. 

Second, we model decoherence as an exponential decay where the probability of no decoherence is $\prod_{k=1}^3 \text{exp}(k*t_k/T_1)$ where $t_k$ is the time the qubit spends in state $k$.  When we construct the circuit we keep track of how long each qudit exists in the $\ket{1}$ or $\ket{3}$ state as the maximum state and calculate the probability of not decohering over the course of the execution for each qudit.  The product of the expected success of each qudit is the EPS due to coherence for the entire circuit.  When multiplied by the gate EPS, we have the EPS for the entire circuit.

\subsection{Circuit Simulation}
Since access to ququart devices at this scale are limited, we must use simulation to evaluate the performance of our approach. We use the trajectory method \cite{brun_simple_2002} for improved scalability compared to full density simulation.  This work simulates circuits of up to 24 qubits (or, equivalently, 12 ququarts). For this work, for each circuit, we generate at least 1000 random quantum states and for each we simulate once and compute the average fidelity over all random states. We emphasize the use of random \textit{quantum} states as classical inputs are not always affected by quantum errors. 

In the past, prior work on simulation of qudit systems neglects the realistic duration differences between gates which results in drastically different usage patterns and simply injecting errors on a moment-to-moment basis can skew results. For example, in this work our direct-to-pulse compilation of CCX and CCZ gates have significantly different execution times. We modify the trajectory method simulation slightly to account for this difference. Rather than inserting many idle gates during each time step, before each gate, we insert one idle gate using the exact time that qudit has been idle.  This is a more accurate representation of from which state these qudits could be decohering.

\begin{figure*}
    \centering
    \scalebox{0.9}{
    \includegraphics[width=\linewidth]{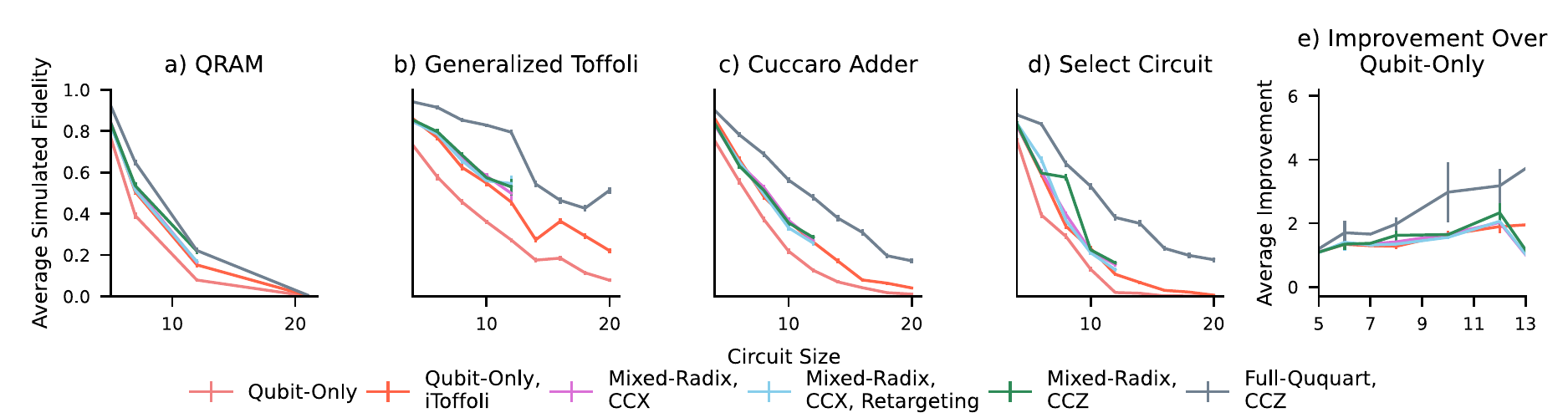}
    }
    \vspace{-0.5em}
    \caption{Simulated results for QRAM, Generalized Toffoli, Cuccaro Adder and Select Circuit from 5 to 21 qubits with different mixed-radix and full-ququart compilation strategies. The mixed-radix strategies do not have complete error bars due to the requirement to simulate a four-level system for every qubit which would require more than 86 GB of memory per circuit in our simulation framework. The final graph is the average fidelity improvement for each compilation method over the qubit-only compilation method as the size of the circuit increases.}
    \label{fig:main-simulation-results}
    \vspace{-0.5em}
\end{figure*}

\subsection{Noise Model for Qudit Systems}
For qubits we consider both symmetric depolarizing and amplitude damping errors. There are four possible single-bit 
channels: no error ($I$), bit flip errors ($X$), phase flip errors ($Z$) and bit and phase flip errors ($Y = ZX$). In simulation each error channel is drawn with probability $p/3$. Two-qubit errors are given as the product of single-qubit errors, e.g. $X \otimes X$ for a bit flip on both interacting qubits; there are 16 possible channels of this type so each error occurs with probability $p/15$ and no error ($I \otimes I$) occurs with probability $1 - 15p$. 

For a general qudit system, we consider a generalized form of these errors. The ``bit-flip" type gates become $X_{+1 \text{mod } d}$ and the ``phase-flip'' errors become $Z_d = \text{diag}(1, \exp{\omega}, \exp{\omega^2}, ..., \\ \exp{\omega^{d-1}})$ where $\omega^j$ is the $j-th$ root of unity. The product of $\{I, X_{+1 \text{mod } d}, ..., X_{+1 \text{mod } d}^{d-1}\}$ and $\{I, Z_d, Z_d^2, ..., Z_d^{d-1}\}$ is a basis for all $d\times d$ Pauli matrices which allows us to construct a general symmetric qudit depolarizing channel. This explains the expected increase in error for using qudit systems: For a two-qubit gate the chance of \textit{no} error is $1 - 15p$ while for a ququart this chance diminishes to $1 - 255p$ let alone possible differences in $p$ \cite{miller_propagation_2018}. 

Amplitude damping for qubits can be described as non-unitary transformations on the quantum state with operators \\ $K_0 = \text{diag}(1, \sqrt{1 - \lambda_1})$ and $K_1 = \sqrt{\lambda_1}e_{0, 1}$. Here $e_{i,j}$ refers to a matrix with all 0's except for a $1$ in the $i$-th row and $j$-th column and is of appropriate dimension. 
In the general qudit case we have \\ $K_0 = \text{diag}(1, \sqrt{1 - \lambda_1}, \sqrt{1 - \lambda_2}, ... \sqrt{1 - \lambda_{d}})$, $K_1 = \sqrt{\lambda_1}e_{0, 1}$, ... $K_d = \sqrt{\lambda_{d}}e_{0, d-1}$. Since we primarily focus on a superconducting system in this study we take $\lambda_m = 1 - \exp{-m\Delta t / T_1}$ where $\Delta t$ is the idling duration and $T_1$ is the coherence time of the qubit \cite{khammassi_qx_2017}. 

In this work we are also concerned with the manipulation of mixed-radix systems. When drawing an error for such a system, for example a qubit-ququart interaction, we consider only relevant errors for the respective participant. For instance, a two-qudit error is drawn from $P_2 \otimes P_4$ and not from $P_4 \otimes P_4$ (where $P_d$ is the set of $d$-dimensional Paulis, exactly the set of potential errors described above). Similarly, for two-qubit gates on encoded qubits, we consider only single \textit{ququart} errors since gates on encoded systems are equivalent to single-ququart gates.
\section{Results}

When we are able to perform native implementations of three-qubit gates via ququarts, we significantly reduce the number of gates that need to be executed, reducing failure rate.  However, the reduction in time to execute these gates may not be enough to overcome the reduced coherence time of higher radix states. In Figure \ref{fig:main-simulation-results}a-d, we examine the simulation fidelities for three-qubit compilation strategies across different sized circuits using Toffoli gate based decompositions.  Each point represents the average fidelity of 1000+ different initial states run once, with randomly inserted error.  The error bars are the standard error, which is the standard deviation of the all the trials divided by the square root of the number of trials.  The mixed-radix compilation schemes stop at 12 qubits due to memory-based computational limitations.  While mixed-radix circuits start in an all-qubit state, we must model them as if they are entirely on ququarts, since we must be able to model the higher levels at all times.  This restricts the number of physical devices we are able to simulate in this scheme to 12 ququarts.

\begin{figure}
    \centering
    \scalebox{0.85}{
    \includegraphics[width=\linewidth]{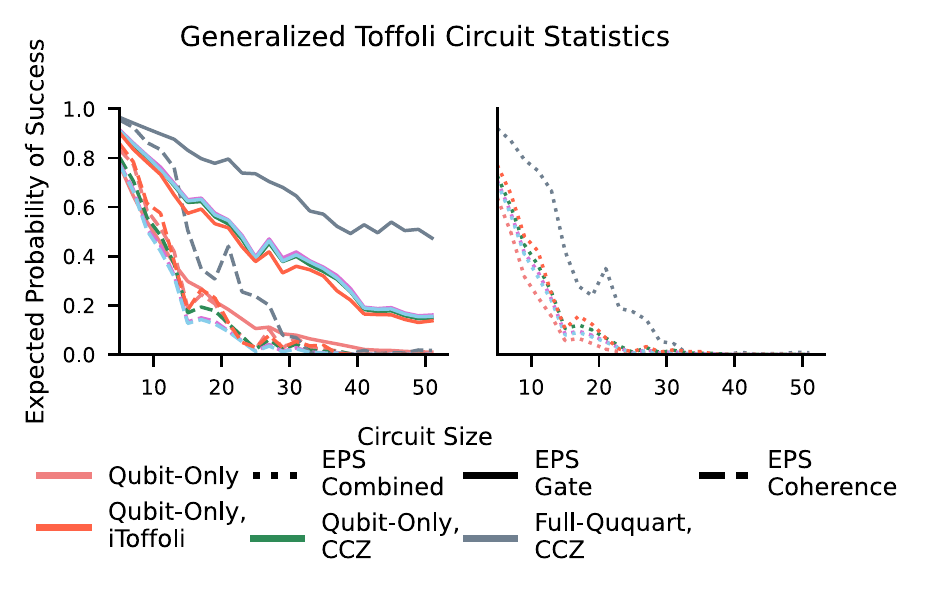}
    }
    \caption{EPS statistics for the generalized Toffoli circuit.  We show the gate and coherence EPS on the left and the product EPS on the right.}
    \label{fig:circuit-estimation}
    \vspace{-1em}
\end{figure}

The first difference is that all of our mixed-radix and full-ququart compilation strategies exceed the fidelity of our baseline two-qubit gate, qubit-only compilation scheme.  From a pure gate error perspective, this should not be unexpected, each of these schemes greatly reduces the number of gates required to execute the same operation. Figure \ref{fig:circuit-estimation} demonstrates how the EPS for gate error is substantially improved by using three two-ququart gates gates or one two-ququart gate for full-ququart computation.  And, as hoped, the simulation finds that the idle time potentially spent in the $\ket{2}$ and $\ket{3}$ state does not outweigh the benefits of the using fewer gates and the shorter circuit duration.  The shorter duration of the gates counteracts the increased decoherence rate of the ququarts. Figure \ref{fig:circuit-estimation} also demonstrates the same point. The coherence EPS between all of the mixed-radix strategies and the qubit-only baseline are nearly the same, and is improved for full-ququart strategies. The general trend of our simulation results is mirrored in Figure \ref{fig:circuit-estimation} where the total EPS of the circuit is shown.  As the EPS trends match what we find in our simulated results, we are able to infer that the scaling of simulation will match the scaling of the EPS results.  While we only show examples of the generalized Toffoli circuit, the results are similar for other circuits.  However, we note that the mixed-radix strategies only marginally outperform or match the simulated results of the qubit-only $i$Toffoli based strategy. This makes sense when we examine the $i$Toffoli based decomposition.  We must insert an extra SWAP gate to perform the corrective Controlled-S gate, resulting in a similar number of gates, and duration, for both decompositions.  Additionally, this SWAP may result in extra corrective SWAPs later on. The extra communication disrupts the layout of the circuit further than was intended, and can require extra gates.

\begin{figure*}
    \centering
    \scalebox{0.85}{
    \includegraphics[width=\linewidth]{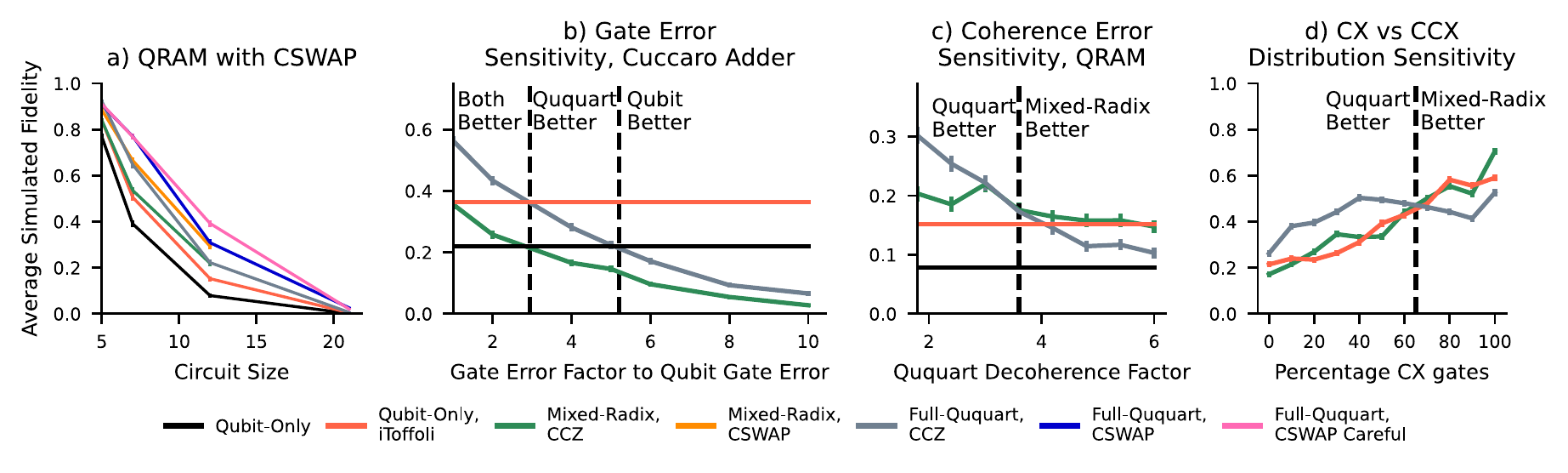}
    }
    \vspace{-0.5em}
    \caption{The results of several sensitivity studies. (a) Sensitivity in simulation by using CSWAP gates in different orientations instead of decomposing to Toffoli gates. (b) Changes in CCZ compilation strategies' fidelities as gate error ququarts increases.  (c) Changes in CCZ compilation strategies' fidelities as coherence error for the $\ket{2}$ and $\ket{3}$ level states changes. (d) Differences in fidelities between mixed-radix and full-ququart compilation strategies as the distribution of CX gates to CCX gates in a circuit changes. In all graphs The black line represents the qubit-only fully-decomposed compilation method.  The red line represents the qubit-only iToffoli-based decomposition. Below those points mixed-radix or full-ququart methods are more error prone than using only qubits. Please note the different scaling on the y-axis.}
    \label{fig:sensitivity-studies}
    \vspace{-0.5em}
\end{figure*}

Digging into the difference between the higher-radix strategies, we find that the mixed-radix strategies are all relatively similar to one another, with some additional separation as the size of the circuit increases.  We first compare the mixed-radix Hadamard corrected CCX gates, shown in light blue, to the mixed-radix gates without this correction, shown in pink.  While there is some cancellation between the single qubit gates, the extra serialization and marginal gate error of the correction gates is a drawback to using this correction strategy based on the simulated results. The reduction in time from the better configuration of the CCX gate is not always enough to overcome these additional costs.  If we instead use CCZ decomposition, shown in green, we consistently achieve the same, or better, fidelity, especially as the size of the circuit increases and the reduction from CCZ gates is more pronounced.  In these cases, the benefits found by using shorter target-independent gates from the start rather than retargeting improves the fidelity of the circuit in this mixed-radix regime.  In Figure \ref{fig:main-simulation-results}e we find that the mixed-radix gates achieve 2x better fidelities for circuit size 12, which is a significant improvement over two-qubit gate computation.  This alone would be a important optimization for three-qubit gate based circuits.

We find that the ququart compilation scheme, shown in grey, has higher fidelity improvement, up to 3x reductions as seen in Figure \ref{fig:main-simulation-results}e, and 50\% improvement over the $i$Toffoli baseline and mixed-radix strategy.  The reasoning behind this is two-fold.  The first is that we no longer need to encode and decode gates before each three-qubit gate in this scheme reducing gate error.  Gate reduction is important, and this reduces the number of gates.  The second is reduction of communication.  With the higher connectivity at all times, we reduce the qubit communication required to perform certain gates. Both of these factors add to the reduction in time, keeping the full-ququart based circuits under the coherence limits and maintaining higher circuit fidelity.  We further reduce the overall circuit time by using faster, target-independent CZ gates in place of CX.

There are cases where full-ququart compilation does not outperform mixed-radix compilation to the same degree.  For instance, the QRAM circuit.  There are more than double the CX gates as Toffolis in this circuit.  The serialization induced by ququarts with slower two-qubit gates reduces the effectiveness of ququarts. Additionally, these benchmarks are only kernels of computations that could be used within the context of larger circuits.  In such cases, we will not have the benefit of a perfect mapping to start.  This would not affect the improvement in fidelity from the qubit only to the mixed-radix strategies, but the effort to encode the qubits into a full-ququart regime before execution may outweigh the benefits.

\subsection{Special Gate Case Study: CSWAP}
As detailed in Section \ref{sec:multitarget} we could instead decompose to a different-three qubit gate in the circuit.  In the case of QRAM, this is the CSWAP gate.  In Figure \ref{fig:sensitivity-studies}a, we explore the differences in fidelity when we use CSWAP gate alongside the original results using CCZ gates.  A CSWAP can be constructed from two CX gates and one CCX gate, but cannot be re-targeted in the same way.  Regardless, in the mixed-radix state, by orienting the CSWAP such that the targets are separate from the controls when possible and like qubits are with like, we see improvements over the CCZ decomposition.  In fact it is able to beat the full-ququart CCZ compilation in some cases because of the reduced number of CX gates.  While we can always attempt to encode the qubits favorably in a mixed-radix environment, this is not as natural a change when compiling for ququarts, and could lead to bad configurations if we solely focus on the disruption of qubits on ququarts.  If we focus on the CSWAP in a full-ququart regime, the basic version shown in blue, and instead use the strategy that places the targets in the same ququart, shown in bright pink, we find even more improvement to our full-ququart encoding regime.  This further indicates the importance using the best decomposition possible by separating the targets and the controls of certain gates.

\subsection{Sensitivity to Ququart Gate Error Rate}
While we synthesized our gates using a realistic Hamiltonian, it is still more difficult to physically realize gates that access higher energy levels.  In Figure \ref{fig:sensitivity-studies}b we explore how the simulated fidelity changes as the error on ququart and mixed-radix gate increases for an 11-qubit Cuccaro Adder.  Both strategies see a very fast drop off as the gate error increases, crossing over the qubit-only baseline fidelity when the ququart error rate is between two and four times worse than qubit gates for mixed-radix compilation (97\% fidelity), and between four and six times worse for full-ququart compilation (94\%).  We also find that the $i$Toffoli strategy outperforms the full-ququart strategy at three times worse ququart gates than qubit gates as well.  While these are still high fidelity targets, it does indicate that we do not need our three-qubit gates exceed the fidelity of two-qubit gates for these strategies to be successful.

\subsection{Sensitivity to Ququart Coherence Error Rate}
In this work we selected the expected theoretical decrease in coherence time as the grounding for most of our simulation experiments.  However, physical realizations don't always meet reality.  Accessing higher energy levels may prove to be more costly in terms of coherence time due to lack of control, or it may be less of an issue as has been found by some testbeds when accessing the qutrit state \cite{cervera-lierta_experimental_2022}.  In Figure \ref{fig:sensitivity-studies}c we demonstrate the effects of changing the rate that the $\ket{2}$ and $\ket{3}$ levels decohere for an 12-qubit QRAM circuit.  The main detail to note is that as the rate increases, the distance between mixed-radix and full-ququart fidelities decreases until mixed-radix becomes higher fidelity.  Mixed-radix gates do not spend as much time in the higher level states, so as machines are developed and these level are more unstable, it may be better to avoid using full ququart encodings for larger circuits.

\subsection{Ratio of Three Qubit Gates to Two Qubit Gates}
It may not always be the case that the number of three qubit gates greatly exceeds the number of two qubit gates. Future applications may have a higher mix of two-qubit gates to three-qubit gates, or may only require a few three qubit gates to perform the desired operation.  In Figure \ref{fig:sensitivity-studies}d, we example how the fidelity of different mixes of two-qubit to three-qubit gates is effected by compilation using a full-ququart strategy versus a mixed-radix strategy for an 11-qubit circuit.  As the ratio of two-qubit to three-qubit gates increases, it becomes less and less profitable to use a full ququart encoding.  At 60\% CX gates it becomes more profitable to remain in the mixed-radix regime. Using CX gates on ququarts requires more serialization, since we cannot perform two separate operations on qubits encoded in the same ququart.  This increases the time, and we start seeing the effects of reduced coherence times.  This changes the calculus about when mixed-radix is better than full-ququart compilation. In cases where we don't need to use as many three qubit gates, it does not make as much sense to use ququarts for the entirety of the circuit.  While this indicates that quantum circuits that only use two-qubit gates do not benefit from this encoding scheme, we can use resynthesis tools \cite{younis_berkeley_2021} to automatically insert three-qubit gates into the circuit, such as in \cite{patel_geyser_2022}.  However, resynthesis can introduce additional error as a perfect direct translation is not always possible and is better explored in a future work. We also include the $i$Toffoli strategy in this analysis as well. We find that it matches the mixed-radix strategy, further solidifying that these strategies have similar performance characteristics.
\section{Related Work}

While this work is the first we are aware of to explore ququart-based execution, there have been studies using existing superconducting qubit technology to execute native three qubit gates.  In particular, Kim et al. \cite{kim_high-fidelity_2022} and Gokhale et al. \cite{gokhale_quantum_2020} have explored driving two connections between three qubits in a line to perform three qubit gates in a superconducting architecture.  Gokhale developed a technique to execute two CX gates in parallel in a single CXX gate.  These gates did not find improved fidelity, but achieved similar goals of faster parallel gate execution than serial execution as described in this work.  It should be noted that this was done without explicit calibration for this sort of operation.  While this work may seem similar through the application of multiqubit gates across many qubits, it mainly focuses on non-superconducting devices, which are able to make use of a global operator gate.  It touches on performing three-qubit gates on superconducting devices, but is unable to generate gates that are more successful than the serialized decomposition. Our work focuses more on superconducting devices and direct synthesis, and must also contend with the issue of communication.

Kim et al. \cite{kim_high-fidelity_2022} developed a 98.2\% fidelity iToffoli gate between three superconducting qubits.  In this case, both controls induce a state change in a center qubit by driving the connections between the qubits.  This gate is performed very quickly with a gate duration of 392 ns.  While an impressive result, it is difficult to compare this work to our own as the device has a substantially different Hamiltonian than this work assumes.  Additionally, this required significant manual calibration between each of the three qubits, a process which may not scale well to larger systems. This work is the basis for our $i$Toffoli baseline, which we found to be similar in performance to the mixed-radix strategy.  However, the computational complexity of generating these pulses is much higher, requiring additional optimizable controls and a larger simulated Hilbert space (when taking into account the simulation of additional ``guard'' energy levels).  Additionally, the mixed-radix scheme presented here only requires calibration between each pair of qudits, which is similar to the processes already in use on quantum computers. The $i$Toffoli scheme would require calibration between each trio of qubits, which would add a significant additional overhead.

There have also been several physical realizations of the $i$Toffoli gate that use the $i$SWAP gate, a CPHASE gate, and a reverse $i$SWAP gate using the $\ket{2}$ state to change the state of a qubit in \cite{hill_realization_2021, fedorov_implementation_2011}. Galda et. al. \cite{galda_implementing_2021} explores using qutrits on IBM's Jakarta device to implement a Toffoli gate with 78\% fidelity.  This is similar to our work, using the more accessible $\ket{2}$ state to perform a three-qubit gate.  These are conceptually similar to the encode, mixed-radix Toffoli, and decode scheme that was laid out in this work. We believe that a machine specifically designed with qudits in mind could enable much higher fidelities for similar experiments.

\section{Conclusion}
The architecturally imposed requirement to decompose more complex three-qubit gates into component one- and two-qubit gates is an extreme hurdle for realizing quantum computing.  Decomposing these gates increases both the number of error-prone gates that need to be executed, and the execution time of the circuit on devices with short coherence times.  However, many architectures have access to higher level states beyond the traditional two-level system.  While more prone to decoherence and error, this extra computational space can be used to compress quantum data, encoding two qubits into one physical device called a ququart.

This work takes advantage of increased connectivity and interaction potential when we have encoded qubits into a four-level system.  Encoding qubits in this way allows for the interaction of three to four qubits across a single physical connection, and we synthesize a library or efficient three-qubit gates via optimal control that take advantage of this virtual connectivity and are much faster and higher fidelity than performing the decomposition of a three-qubit gate.  We also demonstrate the viability of this encoding scheme and gate set via the execution of a $H \otimes H$ gate on real superconducting hardware. We then use these gates to develop compilation strategies, the quantum waltz, that use the most efficient configurations of three-qubit gates on mixed-radix and full-ququart systems to produce circuits that achieve 2x and 3x better simulated fidelities in mixed-radix and full-ququart environments, respectively compared to two-qubit based strategies.  We also demonstrate that ququart-based gates are a viable alternative to $i$Toffoli based three-qubit pulse strategies with potential practical upsides. Despite the difficulty of accessing and performing operations on higher level states, this efficient implementation of three-qubit gates provides worthwhile benefits for quantum computation.

Mixed-radix and full-ququart implementations of three-qubit gates makes ququart computation an invaluable piece of the quantum computing repertoire. It is more flexible than previous hand optimized circuits to improve circuit execution via higher radix devices, does not require the use of quaternary-based logic, and can be selectively applied to certain sections of larger circuits.  Realized implementations of these gates provide a massive opportunity to improve near-term execution of quantum circuits and expand the capabilities of quantum computers.



\begin{acks}
This work is funded in part by EPiQC, an NSF Expedition in Computing, under award CCF-1730449; in part by STAQ under award NSF Phy-1818914; in part by NSF award 2110860; in part by the US Department of Energy Office of Advanced Scientific Computing Research, Accelerated Research for Quantum Computing Program; and in part by the NSF Quantum Leap Challenge Institute for Hybrid Quantum Architectures and Networks (NSF Award 2016136) and in part based upon work supported by the 
U.S. Department of Energy, Office of Science, National Quantum Information Science Research Centers.  FTC is Chief Scientist for Quantum Software at ColdQuanta and an advisor to Quantum Circuits, Inc.

We would like to thank Casey Duckering for his input in early discussion of compiler development for ququarts. We would like to thank Stefanie Günther and N. Anders Petersson for valuable advice on using the quantum optimal control software packages Juqbox and Quandary.

This work was completed in part with resources provided by the University of Chicago’s Research Computing Center.
\end{acks}

\bibliographystyle{ACM-Reference-Format}
\balance
\bibliography{references3}

\end{document}